\documentclass[aps, prl, reprint, superscriptaddress, floatfix, showpacs, amsmath, amssymb]{revtex4-1}

\usepackage{lineno}
\usepackage{graphicx}
\usepackage{dcolumn}
\usepackage{bm}
\usepackage{epstopdf} 
\usepackage[usenames,dvipsnames,pdftex]{color} 
\usepackage{hyperref}

\newcommand {\MM} [1] {\ensuremath{#1}}

\newcommand {\IT} [1] {\ensuremath{#1}}

\newcommand {\SUB} [2] {\MM{#1\ensuremath{_{#2}}}}
\newcommand {\SUP} [2] {\MM{#1\ensuremath{^{#2}}}}

\newcommand {\momentum} {\IT{p}}

\newcommand {\transverse} {\IT{T}}
\newcommand {\pT} {\SUB{\momentum}{\transverse}}

\newcommand {\TIMES} {\ensuremath{\times}}

\newcommand {\electron} {\IT{e}}

\newcommand {\program} [1] {\textsc{#1}}
\newcommand {\PYTHIA} {\program{pythia}}
\newcommand {\GEANT} {\program{geant}}

\newcommand {\capsword} [1] {\textsc{#1}}

\newcommand {\QCD} {\capsword{QCD}}

\newcommand {\DIS} {\capsword{DIS}}
\newcommand {\RHIC} {\capsword{RHIC}}
\newcommand {\PHENIX} {\capsword{PHENIX}}
\newcommand {\STAR} {\capsword{STAR}}

\newcommand {\BEMC} {\capsword{BEMC}}
\newcommand {\EEMC} {\capsword{EEMC}}
\newcommand {\ESMD} {\capsword{ESMD}}
\newcommand {\TPC} {\capsword{TPC}}

\newcommand {\Wvar} {\IT{W}}
\newcommand {\Wpl} {\SUP{\Wvar}{+}}
\newcommand {\Wmi} {\SUP{\Wvar}{-}}

\newcommand {\Wpm} {\SUP{\Wvar}{\pm}}
\newcommand {\epm} {\SUP{\electron}{\pm}}
\newcommand {\taupm} {\SUP{\tau}{\pm}}
\newcommand {\Zgam} {\ensuremath{\IT{Z}/\gamma^*}}

\newcommand {\Wpmtoenu}{\ensuremath{\Wpm \to \epm \nu_e}}
\newcommand {\Wpmtaunu} {\ensuremath{\Wpm \to \taupm \nu_\tau}}
\newcommand {\Zgtoee} {\ensuremath{\Zgam \to e^+e^-}}

\newcommand {\EeT} {\ensuremath{E^\electron_T}}

\newcommand {\RHICBOS} {\program{rhicbos}}
\newcommand {\CHE} {\program{che}}
\newcommand {\TAUOLA} {\program{tauola}}
\newcommand {\mee}  {\ensuremath{m_{e^+e^-}}}

\begin{document}

\title{\texorpdfstring{Measurement of Longitudinal Spin Asymmetries for Weak Boson Production in Polarized Proton-Proton Collisions at RHIC}{Title}}

\affiliation{AGH University of Science and Technology, Cracow, Poland}
\affiliation{Argonne National Laboratory, Argonne, Illinois 60439, USA}
\affiliation{University of Birmingham, Birmingham, United Kingdom}
\affiliation{Brookhaven National Laboratory, Upton, New York 11973, USA}
\affiliation{University of California, Berkeley, California 94720, USA}
\affiliation{University of California, Davis, California 95616, USA}
\affiliation{University of California, Los Angeles, California 90095, USA}
\affiliation{Universidade Estadual de Campinas, Sao Paulo, Brazil}
\affiliation{Central China Normal University (HZNU), Wuhan 430079, China}
\affiliation{University of Illinois at Chicago, Chicago, Illinois 60607, USA}
\affiliation{Cracow University of Technology, Cracow, Poland}
\affiliation{Creighton University, Omaha, Nebraska 68178, USA}
\affiliation{Czech Technical University in Prague, FNSPE, Prague, 115 19, Czech Republic}
\affiliation{Nuclear Physics Institute AS CR, 250 68 \v{R}e\v{z}/Prague, Czech Republic}
\affiliation{Frankfurt Institute for Advanced Studies FIAS, Germany}
\affiliation{Institute of Physics, Bhubaneswar 751005, India}
\affiliation{Indian Institute of Technology, Mumbai, India}
\affiliation{Indiana University, Bloomington, Indiana 47408, USA}
\affiliation{Alikhanov Institute for Theoretical and Experimental Physics, Moscow, Russia}
\affiliation{University of Jammu, Jammu 180001, India}
\affiliation{Joint Institute for Nuclear Research, Dubna, 141 980, Russia}
\affiliation{Kent State University, Kent, Ohio 44242, USA}
\affiliation{University of Kentucky, Lexington, Kentucky, 40506-0055, USA}
\affiliation{Korea Institute of Science and Technology Information, Daejeon, Korea}
\affiliation{Institute of Modern Physics, Lanzhou, China}
\affiliation{Lawrence Berkeley National Laboratory, Berkeley, California 94720, USA}
\affiliation{Massachusetts Institute of Technology, Cambridge, Massachusetts 02139-4307, USA}
\affiliation{Max-Planck-Institut f\"ur Physik, Munich, Germany}
\affiliation{Michigan State University, East Lansing, Michigan 48824, USA}
\affiliation{Moscow Engineering Physics Institute, Moscow Russia}
\affiliation{National Institute of Science Education and Research, Bhubaneswar 751005, India}
\affiliation{Ohio State University, Columbus, Ohio 43210, USA}
\affiliation{Old Dominion University, Norfolk, Virginia 23529, USA}
\affiliation{Institute of Nuclear Physics PAN, Cracow, Poland}
\affiliation{Panjab University, Chandigarh 160014, India}
\affiliation{Pennsylvania State University, University Park, Pennsylvania 16802, USA}
\affiliation{Institute of High Energy Physics, Protvino, Russia}
\affiliation{Purdue University, West Lafayette, Indiana 47907, USA}
\affiliation{Pusan National University, Pusan, Republic of Korea}
\affiliation{University of Rajasthan, Jaipur 302004, India}
\affiliation{Rice University, Houston, Texas 77251, USA}
\affiliation{University of Science and Technology of China, Hefei 230026, China}
\affiliation{Shandong University, Jinan, Shandong 250100, China}
\affiliation{Shanghai Institute of Applied Physics, Shanghai 201800, China}
\affiliation{SUBATECH, Nantes, France}
\affiliation{Temple University, Philadelphia, Pennsylvania 19122, USA}
\affiliation{Texas A\&M University, College Station, Texas 77843, USA}
\affiliation{University of Texas, Austin, Texas 78712, USA}
\affiliation{University of Houston, Houston, Texas 77204, USA}
\affiliation{Tsinghua University, Beijing 100084, China}
\affiliation{United States Naval Academy, Annapolis, Maryland, 21402, USA}
\affiliation{Valparaiso University, Valparaiso, Indiana 46383, USA}
\affiliation{Variable Energy Cyclotron Centre, Kolkata 700064, India}
\affiliation{Warsaw University of Technology, Warsaw, Poland}
\affiliation{University of Washington, Seattle, Washington 98195, USA}
\affiliation{Wayne State University, Detroit, Michigan 48201, USA}
\affiliation{Yale University, New Haven, Connecticut 06520, USA}
\affiliation{University of Zagreb, Zagreb, HR-10002, Croatia}

\author{L.~Adamczyk}\affiliation{AGH University of Science and Technology, Cracow, Poland}
\author{J.~K.~Adkins}\affiliation{University of Kentucky, Lexington, Kentucky, 40506-0055, USA}
\author{G.~Agakishiev}\affiliation{Joint Institute for Nuclear Research, Dubna, 141 980, Russia}
\author{M.~M.~Aggarwal}\affiliation{Panjab University, Chandigarh 160014, India}
\author{Z.~Ahammed}\affiliation{Variable Energy Cyclotron Centre, Kolkata 700064, India}
\author{I.~Alekseev}\affiliation{Alikhanov Institute for Theoretical and Experimental Physics, Moscow, Russia}
\author{J.~Alford}\affiliation{Kent State University, Kent, Ohio 44242, USA}
\author{C.~D.~Anson}\affiliation{Ohio State University, Columbus, Ohio 43210, USA}
\author{A.~Aparin}\affiliation{Joint Institute for Nuclear Research, Dubna, 141 980, Russia}
\author{D.~Arkhipkin}\affiliation{Brookhaven National Laboratory, Upton, New York 11973, USA}
% ...

\collaboration{STAR Collaboration}\noaffiliation


\pacs{24.85.+p, 13.38.Be, 13.38.Dg, 14.20.Dh}


\begin{abstract} 

We report measurements of single- and double-spin asymmetries for $W^{\pm}$ and $Z/\gamma^*$ boson production in longitudinally polarized $p+p$ collisions at $\sqrt{s} = 510$ GeV by the STAR experiment at RHIC. The asymmetries for $W^{\pm}$ were measured as a function of the decay lepton pseudorapidity, which provides a theoretically clean probe of the proton's polarized quark distributions at the scale of the $W$ mass. The results are compared to theoretical predictions, constrained by polarized deep inelastic scattering measurements, and show a preference for a sizable, positive up antiquark polarization in the range $0.05<x<0.2$.

\end{abstract}

\maketitle 

In high-energy proton-proton collisions, weak boson and Drell-Yan production are dominated by quark-antiquark annihilations.  Because of the valence quark structure of the proton, these interactions primarily involve the lightest two quark flavors, up ($u$) and down ($d$).  In unpolarized collisions, measurements of these processes are used to constrain the helicity-independent parton distribution functions (PDFs) of the quarks ({\it e.g.} Refs.~\cite{Lai:2010vv,Martin:2009iq}).  In particular, Drell-Yan measurements~\cite{Baldit:1994jk,Towell:2001nh} and earlier deep inelastic scattering (DIS) results~\cite{Amaudruz:1991at,Arneodo:1994sh} have reported a large enhancement in $\bar{d}$ over $\bar{u}$ quarks for a wide range of partonic momentum fractions $x$.  Calculations have shown that perturbative QCD does not produce such a flavor asymmetry in the proton's light antiquark distributions, indicating another, likely nonperturbative, mechanism is needed~\cite{Ross:1978xk,Steffens:1996bc}.  This generated significant theoretical interest, with many nonperturbative models able to qualitatively describe the data~\cite{Garvey:2001yq,Kumano:2001cu,Dressler:1999zv,Bourrely:2001du,Bourrely:2013qfa}.
nonperturbative
In the case of longitudinally polarized proton collisions at \RHIC{}, the coupling of \Wpm{} bosons to left-handed quarks and right-handed antiquarks ($u_L\bar{d}_R{\rightarrow}\Wpl$ and $d_L\bar{u}_R{\rightarrow}\Wmi$) determines the helicity of the incident quarks.  This provides a direct probe of the helicity-dependent PDFs through a parity-violating longitudinal single-spin asymmetry which is defined as $A_L~=~(\sigma_+{-}\sigma_-)/(\sigma_+{+}\sigma_-)$, where $\sigma_{+(-)}$ is the cross section when the polarized proton beam has positive (negative) helicity.  Analogous to the unpolarized case, measurements of this asymmetry can be used to constrain the helicity-dependent quark PDFs $\Delta q = q^+ - q^-$, where $q^+(q^-)$ is the distribution of quarks with spin parallel (antiparallel) to the proton spin.  Of particular interest is a possible flavor asymmetry in the polarized case, given by $\Delta\bar{u}{-}\Delta\bar{d}$, which some nonperturbative models predict to be similar to, or even larger than, the unpolarized flavor asymmetry~\cite{Dressler:1999zv,Bourrely:2001du,*Bourrely:2013qfa}.  

Semi-inclusive \DIS{} measurements with polarized beams and targets also constrain the helicity-dependent PDFs, although they require the use of fragmentation functions to relate the measured final-state hadrons to the flavor-separated quark and antiquark distributions~\cite{Adeva:1997qz,Airapetian:2004zf,Alekseev:2010ub}.  Both inclusive and semi-inclusive \DIS{} measurements have been included in global \QCD{} analyses to determine the helicity-dependent PDFs of the proton~\cite{deFlorian:2008mr,*deFlorian:2009vb,*deFlorian:2009vb,Leader:2010rb}.  The extracted polarized flavor asymmetry $\Delta\bar{u}{-}\Delta\bar{d}$ is positive within the sizable uncertainty afforded by the current measurements.

In this Letter, we report measurements of single- and double-spin asymmetries for weak boson production in longitudinally polarized $p{+}p$ collisions from 2011 and 2012 by the \STAR{} collaboration at \RHIC{} for $\sqrt{s}=500$ and $510$~GeV, respectively.  The beam polarization and luminosity of this data set correspond to an order of magnitude reduction in the statistical variance for single-spin asymmetry measurements, in comparison to results reported previously by \STAR~\cite{Aggarwal:2010vc} and \PHENIX~\cite{Adare:2010xa}.  These measurements place new constraints on the helicity-dependent antiquark PDFs, and prefer a larger value for the up antiquark polarization $\Delta\bar{u}$ than previously expected by global QCD analyses~\cite{deFlorian:2008mr,*deFlorian:2009vb,Leader:2010rb}.  

The polarizations of the two beams were each measured using Coulomb-nuclear interference proton-carbon polarimeters, which were calibrated with a polarized hydrogen gas-jet target~\cite{Polarimetry}.  The average luminosity-weighted beam polarization during 2011 (2012) was 0.49~(0.56), with a relative scale uncertainty of 3.4\% for the single beam polarization and 6.5\% for the product of the polarizations from two beams.  The integrated luminosities of the data sets from 2011 and 2012 are 9 and 77~pb$^{-1}$, respectively.

The subsystems of the \STAR{} detector~\cite{Ackermann:2002ad} used in this measurement are the Time Projection Chamber~\cite{Anderson:2003ur} (\TPC), providing charged particle tracking for pseudorapidity $|\eta|\lesssim1.3$, and the Barrel~\cite{Beddo:2002zx} and Endcap~\cite{Allgower:2002zy} Electromagnetic Calorimeters (\BEMC, \EEMC).  These lead-sampling calorimeters cover the full azimuthal angle $\phi$ for $|\eta|<1$ and $1.1<\eta<2$, respectively.  

In this analysis, \Wpm~bosons were detected via their \Wpmtoenu~decay channels, and were recorded using a calorimeter trigger requirement of 12~(10)~GeV of transverse energy $E_T$ in a $\Delta\eta \TIMES \Delta\phi$ region of ${\sim}0.1 \TIMES 0.1$ of the \BEMC~(\EEMC).  Primary vertices were reconstructed along the beam axis of the \TPC~within $\pm100$~cm of the center of the STAR interaction region. The vertex distribution was approximately Gaussian with an rms of 49~cm.  The spread of the vertex distribution allows the detector $\eta$ coverage to be extended by ${\sim}0.1$.

The selection criteria for electrons and positrons detected in the \BEMC{}, with \epm{} pseudorapidity $|\eta_e|<1.1$, are described in previously reported measurements of the $\Wpm$ and $\Zgam$ cross sections~\cite{Adamczyk:2011aa}, and will only be summarized here.  At midrapidity, \Wpmtoenu~events are characterized by an isolated $\epm$ with a transverse energy $\EeT$ measured in the \BEMC{} that peaks near half the $\Wvar$ boson mass.  Leptonic $\Wpm$ decays also produce a neutrino, close to opposite in azimuth of the decay $\epm$.  The neutrino is undetected and leads to a large missing transverse energy.  As a result, there is a large imbalance in the vector transverse momentum ($\pT$) sum of all reconstructed final-state objects for $\Wpm$ events, in contrast to $\Zgtoee$ and QCD dijet events.  We define a $\pT$-balance variable $\vec{p}_T^{~bal}$ which is the vector sum of the \epm~candidate $\vec{p}_T^{~e}$ and the \pT~vectors of all reconstructed jets outside an isolation cone around the \epm~candidate track with a radius of $\Delta R=\sqrt{\Delta\eta^2 + \Delta\phi^2} = 0.7$.  Jets were reconstructed from charged tracks in the \TPC{} and energy deposits in the \BEMC{} and \EEMC{} using an anti-$k_T$ algorithm~\cite{Cacciari:2008gp}.  The scalar variable signed~$\pT$-balance = $(\vec{p}_T^{~bal} \cdot \vec{p}_T^{~e})/|\vec{p}_T^{~e}|$ is required to be larger than 14~GeV$/c$.

\begin{figure}[t] 
	\includegraphics[width=0.483\textwidth]{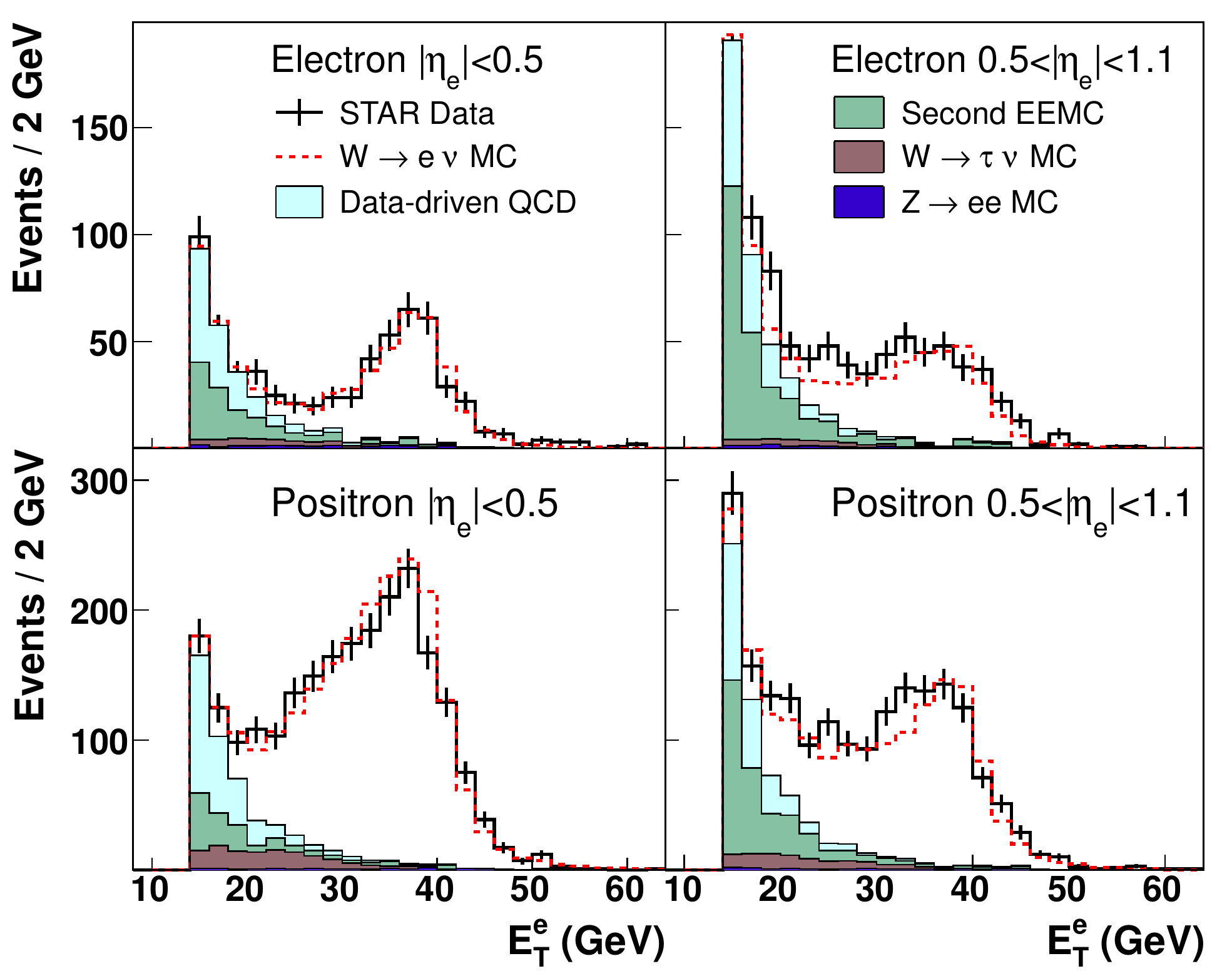}
    	\caption{(color online) $E_T^e$ distribution of  $W^-$ (top) and $W^+$ (bottom) candidate events (black), background contributions, and sum of backgrounds and $\Wpmtoenu$ MC signal (red dashed).}
    	\label{Fig:BEMCW}
\end{figure}

\Wpm~candidates were charge separated based on \epm~track curvature measured in the \TPC.  The charge separated yields are shown in Fig.~\ref{Fig:BEMCW}, along with the estimated contributions from electroweak processes and QCD backgrounds, as a function of $E_T^e$.  The $\Wpmtaunu$ and $\Zgtoee$ electroweak contributions were determined from Monte Carlo (MC) samples simulated using \PYTHIA~6.422~\cite{Sjostrand:2006za} with the Perugia~0 tune~\cite{Skands:2010ak}.  The generated events were passed through a \GEANT~\cite{Brun:1978fy} model of the STAR detector response, embedded in real STAR zero-bias triggered events~\cite{Adamczyk:2011aa}, and reconstructed using the same selection criteria as the data.  In the $\Wpmtaunu$ sample the \TAUOLA{} package was used for the polarized $\taupm$ decay~\cite{Golonka:2003xt}.  Background yields from QCD processes were estimated independently for each $\eta_e$ bin through two contributions described in Ref.~\cite{Adamczyk:2011aa}, referred to as the second \EEMC{} and data-driven QCD.  These background contributions originate primarily from events thatmidrapidity satisfy the $\Wpm$ selection criteria but contain jets escaping detection due to the missing calorimeter coverage for $\eta < -1$ and $\eta>2$.  

\begin{figure}[t] 
	\includegraphics[width=0.483\textwidth]{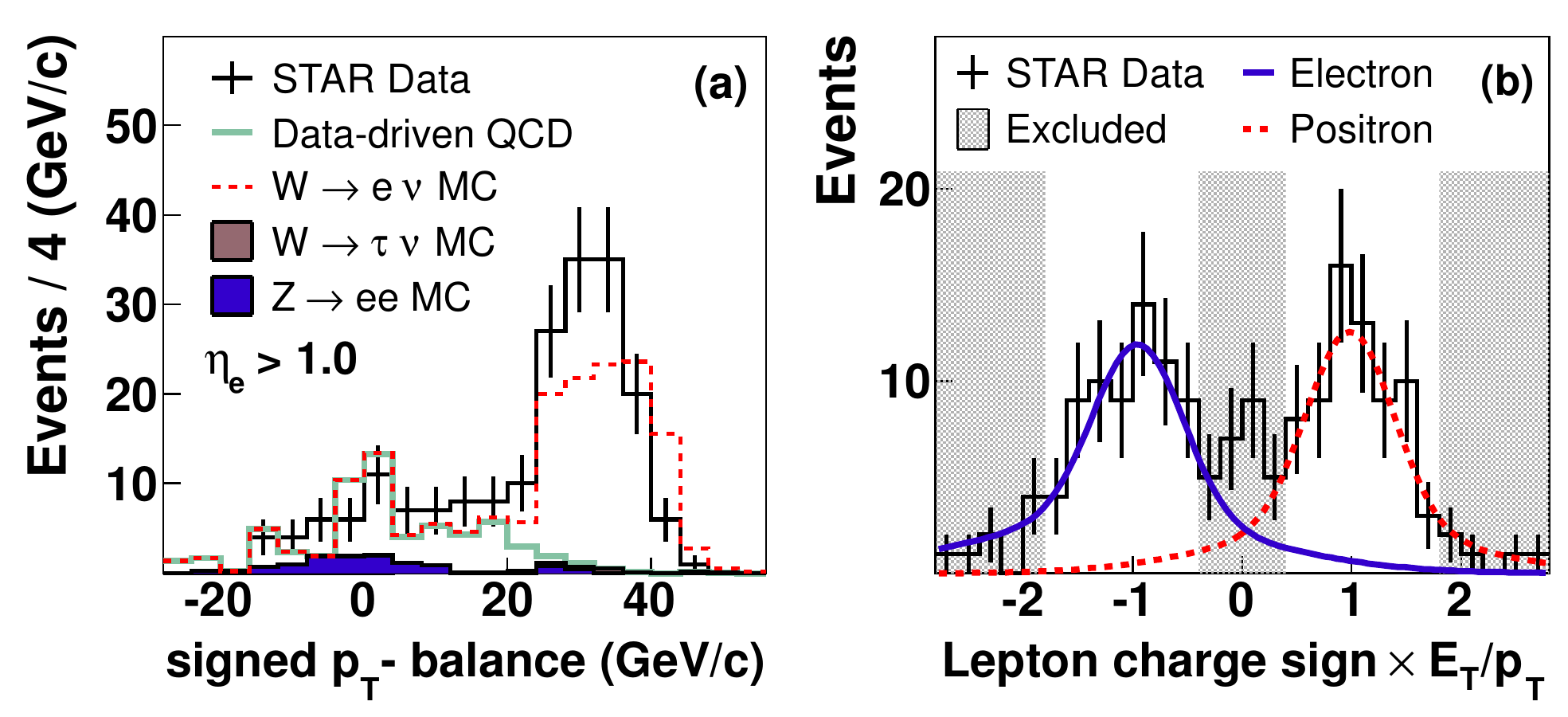} 
	\caption{(color online) (a) Signed $p_{T}$-balance distribution for \epm~candidates reconstructed in the EEMC and (b) distribution of the product of the TPC reconstructed charge sign and $E_T/p_T$.}
    	\label{Fig:EEMCW}
\end{figure}

The EEMC was used to reconstruct the energy of the decay \epm{} candidates at forward rapidity ($\eta_e~>~1$).   Charged track reconstruction was provided by the TPC, limiting the pseudorapidity acceptance to $\eta_e~\lesssim~1.3$.  Similar to the midrapidity event selection, isolation and signed~$\pT$-balance requirements were used to select $\Wpmtoenu$ candidates.  Additionally, the \EEMC~Shower Maximum Detector (\ESMD)~\cite{Allgower:2002zy}, consisting of two orthogonal planes of scintillating strips at a depth of $\sim$5 radiation lengths, provided a measurement of the electromagnetic shower's profile transverse to its propagation direction.  A single electromagnetic shower from a \Wpmtoenu~decay should be isolated with a narrow transverse profile (Moli\`{e}re radius of $\sim$1.5~cm in lead~\cite{Beringer:1900zz}), while QCD background candidates typically contain a $\pi^0$ or other additional energy deposits in proximity to the candidate track leading to a wider reconstructed shower.  In addition, the location of the extrapolated TPC~track and the shower reconstructed in the \ESMD~should be well correlated for \Wpmtoenu~events.  To further suppress QCD background events, a ratio of the energy deposited in the \ESMD{} strips within $\pm 1.5$~cm of the candidate \TPC~track to the energy deposited in the strips within $\pm 10$~cm was computed.  This ratio, denoted $R_{\textrm{ESMD}}$, was required to be greater than 0.6 to select isolated, narrow $\epm$ showers.

The charge-summed candidate yield as a function of signed $p_{T}$-balance for forward rapidity $\epm$ is shown in Fig.~\ref{Fig:EEMCW}(a), where the electroweak contributions were estimated using the same MC samples described for the midrapidity case.  The QCD background was estimated from the shape of the signed $p_T$-balance distribution for $\epm$ candidates with $R_{\textrm{ESMD}}~<~0.5$.  This shape was determined for each charge sign independently and was normalized to the measured yield in the QCD background dominated region, \mbox{$-8 < $ signed $p_T$-balance $ < 8$~GeV$/c$}.   Forward rapidity $\Wpm$ candidates were selected by requiring signed $p_T$-balance $>$ 20~GeV$/c$.  The difference between the data and \Wpmtoenu~MC distributions for signed $p_T$-balance $ > 20$~GeV$/c$ is within the MC normalization uncertainty, and this uncertainty provides a negligible contribution to the measured spin asymmetries.

Figure~\ref{Fig:EEMCW}(b) shows the reconstructed charge sign multiplied by the ratio of $\EeT$ (measured by the EEMC) to $p_T^e$ (measured by the TPC) for forward rapidity candidates.  Because of their forward angle, these tracks have a reduced number of points along their trajectory measured by the TPC compared to the midrapidity case, which leads to a degraded $p_T$ resolution.  Despite that, a clear charge sign separation is observed.  The data were fit to two double-Gaussian template shapes generated from $\Wpm$ MC samples to estimate the reconstructed charge sign purity.  The shaded regions were excluded from the analysis to remove tracks with poorly reconstructed $p_T$ and reduce the opposite charge sign contamination.  The residual charge sign contamination is estimated to be 6.5\%, which is small relative to the statistical uncertainties of the measured spin asymmetries.

\begin{figure} 
	\includegraphics[width=0.483\textwidth]{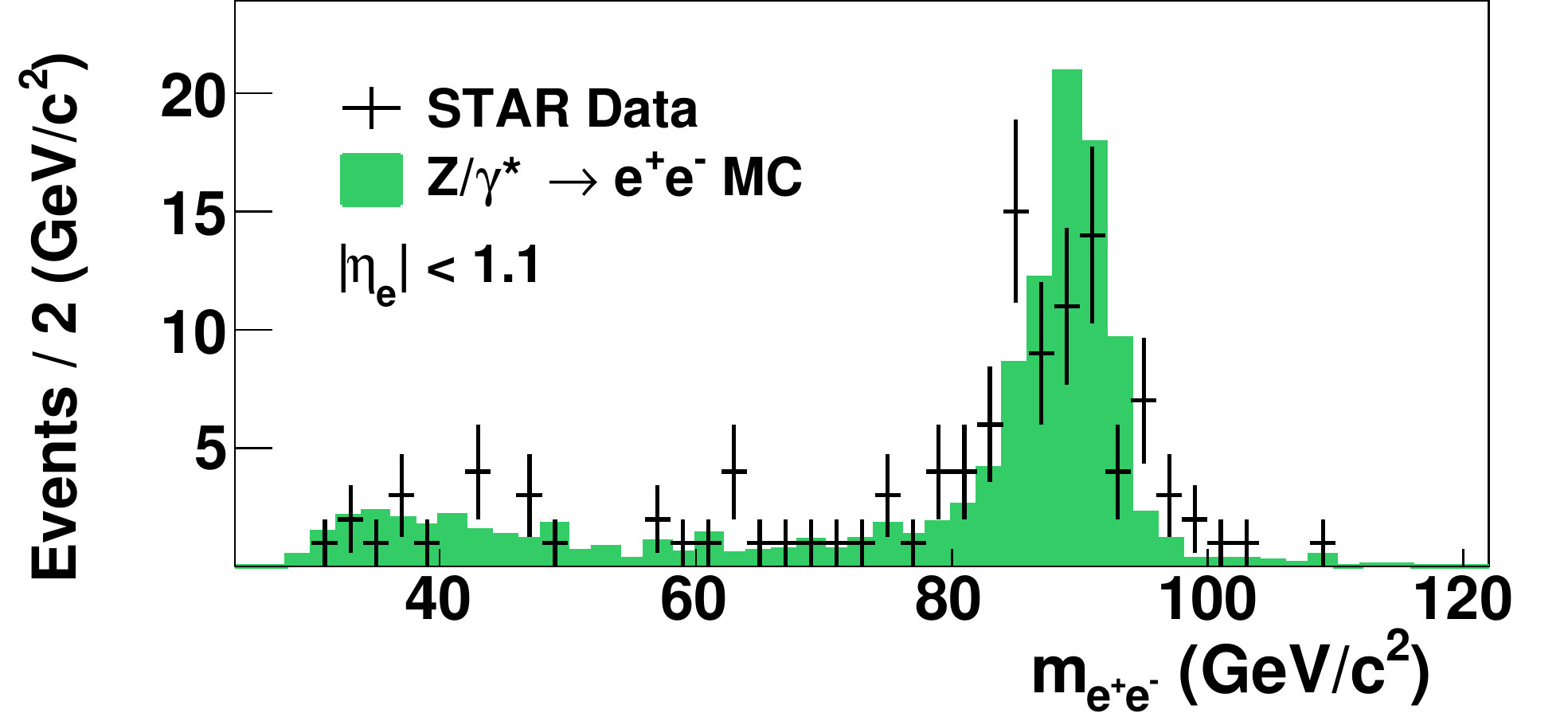} 
    	\caption{(color online) Distributions of the invariant mass of $\Zgtoee$ candidate events. The $\Zgtoee$ MC distribution (filled histogram) is shown for comparison.}
    	\label{Fig:Z}
\end{figure}

Measurements of $\Zgam$ production at RHIC energies are limited by a small production cross section.  However, one unique advantage of this channel is the fully reconstructed $e^+ e^-$ final state, allowing the initial state kinematics to be determined event by event at leading order.  A sample of 88 $\Zgtoee$ events was identified by selecting a pair of isolated, oppositely charged $\epm$ candidates, as described in Ref.~\cite{Adamczyk:2011aa}.  The resulting invariant mass distribution of $e^+ e^-$ pairs is shown in Fig.~\ref{Fig:Z}, superimposed with the MC expectation.  

The measured spin asymmetries were obtained from the 2011 and 2012 data samples using a likelihood method to treat the low statistics of the 2011 sample.  For a given data sample, a model for the expected, spin-dependent $\Wpm$ event yield $\mu$ in a given positive pseudorapidity range, labeled $a$, of the \STAR~detector can be defined for each of the four \RHIC~helicity states of the two polarized proton beams
\begin{equation} 
\begin{array}{l}
\mu_{++}^a = l_{++}N^a ( 1 + P_1 \beta A_L^{+\eta_e} + P_2 \beta A_{L}^{-\eta_e} + P_1 P_2 \beta A_{LL} )\\
\mu_{+-}^a = l_{+-}N^a ( 1 + P_1 \beta A_L^{+\eta_e} - P_2 \beta A_{L}^{-\eta_e} - P_1 P_2 \beta A_{LL} )\\
\mu_{-+}^a = l_{-+}N^a ( 1 - P_1 \beta A_L^{+\eta_e} + P_2 \beta A_{L}^{-\eta_e} - P_1 P_2 \beta A_{LL} )\\
\mu_{--}^a = l_{--}N^a ( 1 - P_1 \beta A_L^{+\eta_e} - P_2 \beta A_{L}^{-\eta_e} + P_1 P_2 \beta A_{LL} )
\label{Eqn:PolModel}
\end{array}
\end{equation} 
where 
\begin{itemize}\addtolength{\itemsep}{-0.5\baselineskip} 
	\item $P_1 (P_2)$ is the absolute value of the polarization of beam 1(2),
	\item $A_L^{+\eta_e}$ ($A_L^{-\eta_e}$) is the single-spin asymmetry measured at positive(negative) $\eta_e$ with respect to beam 1, 
	\item $A_{LL}$ is the parity-conserving double-spin asymmetry~\footnote{The double-spin asymmetry is defined as \begin{equation} A_{LL} = \frac{(\sigma_{++} + \sigma_{--}) - (\sigma_{+-} + \sigma_{-+})}{(\sigma_{++} + \sigma_{--}) + (\sigma_{+-} + \sigma_{-+})} \end{equation} where $\sigma_{+-}$ represents a cross section for beam protons with helicity $(+)$ and $(-)$.} which is symmetric with respect to $\eta_e$,
	\item $N^a$ is the spin averaged yield, and 
	\item $l_{\pm\pm}$ are the respective relative luminosities determined from an independent sample of QCD events, which required a nonisolated lepton candidate with $\EeT<20$~GeV as described in Ref.~\cite{Aggarwal:2010vc}.
\end{itemize}

A similar set of four equations can be written for the symmetric negative pseudorapidity range of the \STAR~detector, labeled $b$, by interchanging $A_L^{+\eta_e}$ with $A_L^{-\eta_e}$.  The dilution of the asymmetries due to unpolarized background contributions to the $\Wpm$ candidate yield are represented by $\beta = S/(S+B)$, where $S$ and $B$ are the number of signal and background events as shown in Figs.~\ref{Fig:BEMCW}~and~\ref{Fig:EEMCW}, and were measured separately for regions $a$ and $b$.  The estimated $\Wpmtaunu$ yield is not a background for the asymmetry measurement as it is produced in the same partonic processes as the primary signal, $\Wpmtoenu$.

The eight spin-dependent yields for the pair of symmetric pseudorapidity regions in the \STAR~detector ($a$ and $b$) are used to define a likelihood function 
\begin{equation} 
	L = \prod_{i}^4 \mathcal{P}(M_{i}^{a} | \mu_{i}^{a}) \mathcal{P}(M_{i}^{b} | \mu_{i}^{b}) g(\beta^a) g(\beta^b)
	\label{Eqn:Likelihood}
\end{equation}
consisting of a product of Poisson probabilities $\mathcal{P}(M_i | \mu_i)$ for measuring $M_i$ events in a helicity configuration, $i$, given the expected value $\mu_i$ from Eqn.~(\ref{Eqn:PolModel}) and a Gaussian probability $g(\beta)$ for the estimated background dilution.  The spin asymmetry parameters ($A_L^{+\eta_e}$, $A_L^{-\eta_e}$ and $A_{LL}$) of this likelihood function were bounded to be within their physically allowed range of [-1,1], $N^{a,b}$ and $\beta^{a,b}$ were treated as nuisance parameters, and the remaining parameters ($P$ and $l_{\pm\pm}$) are known constants.

\begin{figure}[t] 
    	\includegraphics[width=0.483\textwidth]{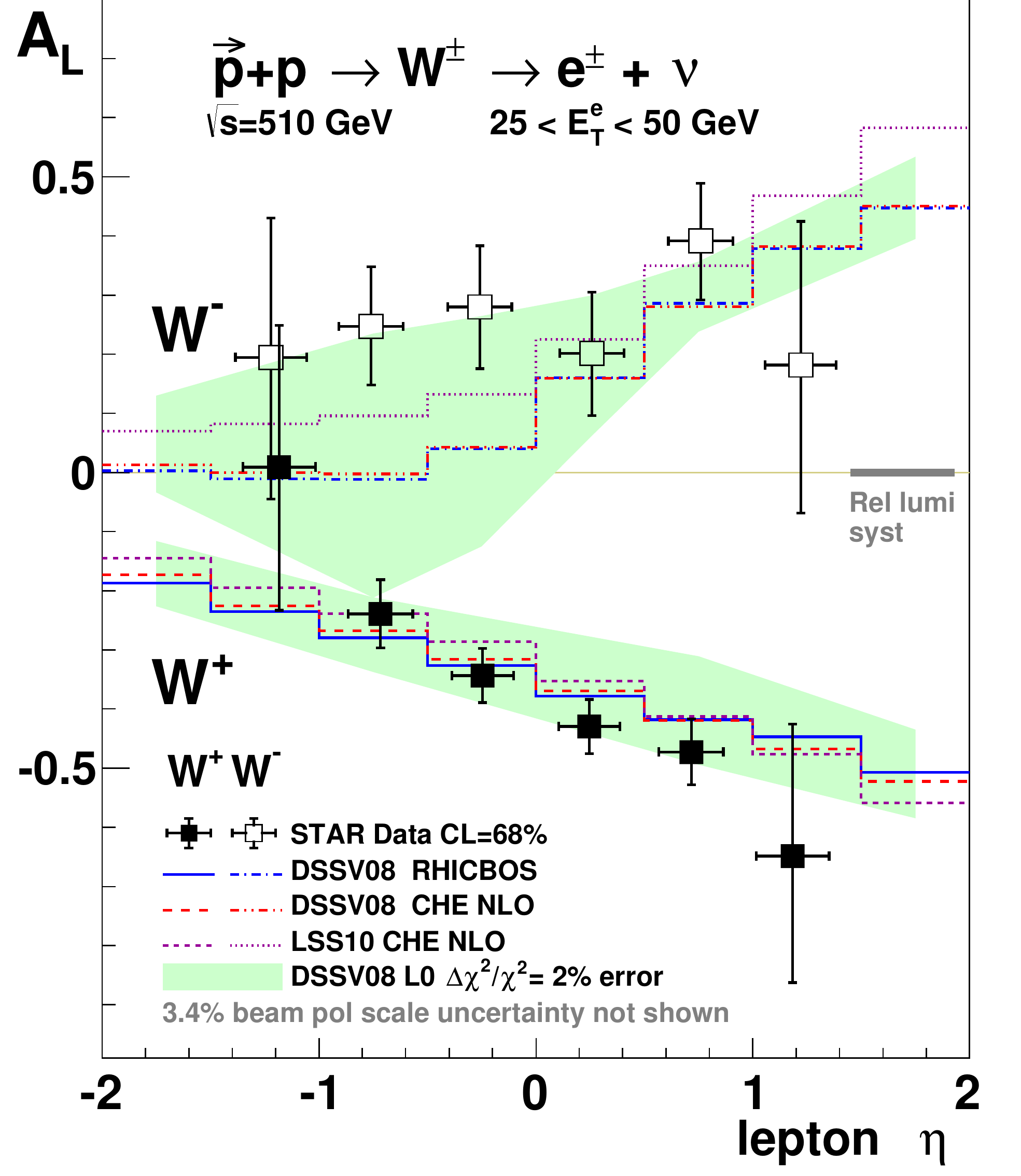} 
	\caption{(color online) Longitudinal single-spin asymmetry $A_L$ for $W^\pm$ production as a function of lepton pseudorapidity $\eta_e$ in comparison to theory predictions (see text for details).}
    	\label{Fig:AL}
\end{figure}

\begin{figure}[t] 
	\includegraphics[width=0.483\textwidth]{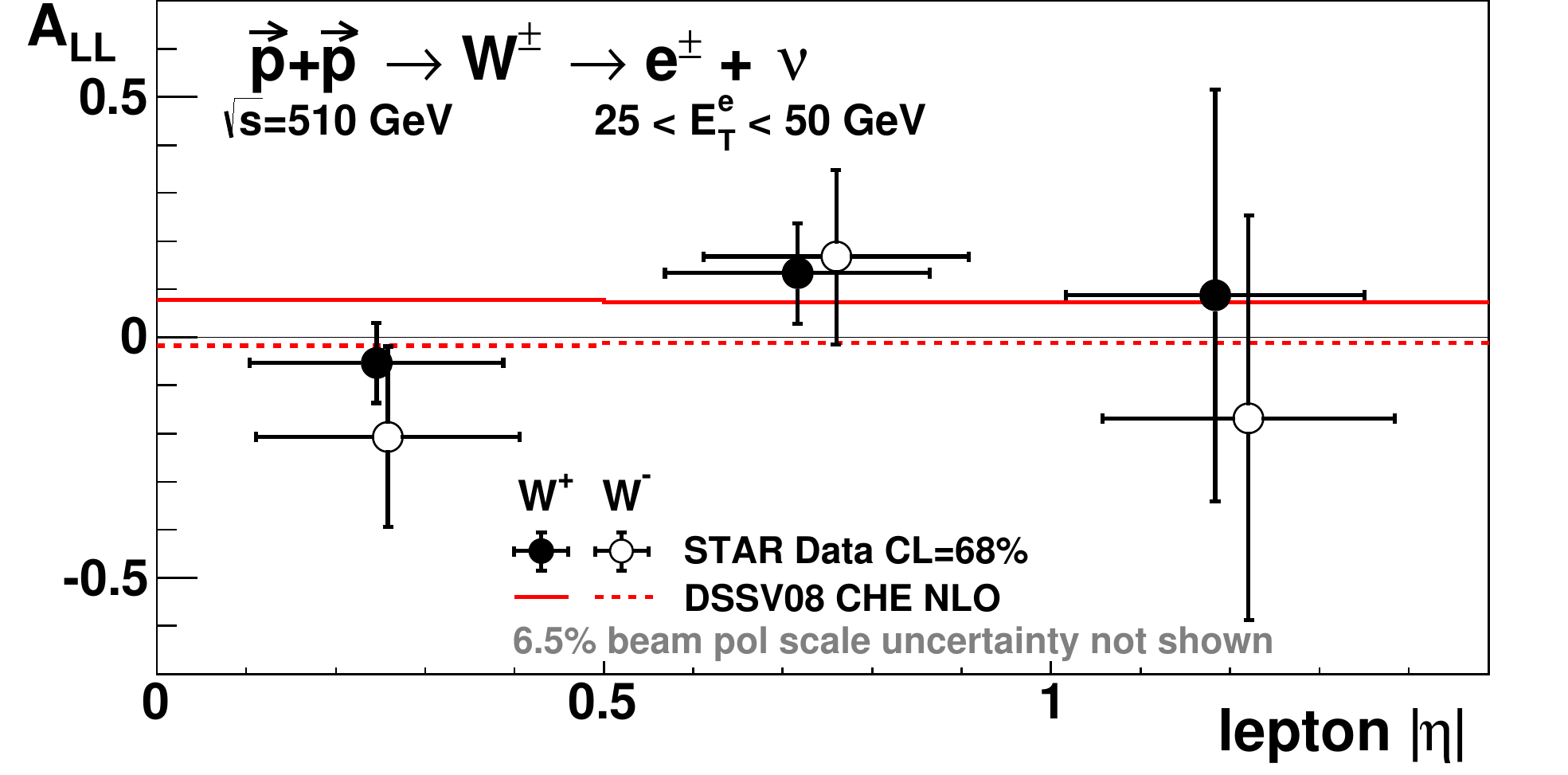} 
	\caption{(color online) Longitudinal double-spin asymmetry $A_{LL}$ for $W^\pm$ production as a function of lepton pseudorapidity $|\eta_e|$ in comparison to theory predictions (see text for details).}
	\label{Fig:ALL}
\end{figure}

Separate likelihood functions were computed for the 2011 and 2012 data sets, consisting of 2759 \Wpl{} and 837 \Wmi{} candidates in total.  The product of these two likelihood functions was used in a profile likelihood analysis~\cite{Beringer:1900zz} to obtain the central values and confidence intervals for the asymmetries.  The \Wpm{} asymmetries were measured for $\epm$ with $25<\EeT<50$~GeV and are shown in Figs.~\ref{Fig:AL} and \ref{Fig:ALL} as a function of $\epm$ pseudorapidity for the single- and double-spin asymmetries, respectively.  These results are consistent with our previous measurements of $A_L$~\cite{Aggarwal:2010vc}.  The data points are located at the average $\eta_e$ within each bin, and the horizontal error bars represent the rms of the $\eta_e$ distribution within that bin.  The vertical error bars show the 68\% confidence intervals, which include the statistical uncertainty, as well as systematic uncertainties due to the unpolarized background dilutions.  The magnitude of the confidence intervals is dominated by the statistical precision of the data.  The relative luminosity systematic uncertainty is $\pm 0.007$ as indicated by the gray band in Fig.~\ref{Fig:AL}.  The single- (double-) spin asymmetries have a common 3.4\% (6.5\%) normalization uncertainty due to the uncertainty in the measured beam polarization.

The measured single-spin asymmetries are compared to theoretical predictions using both next-to-leading order (\CHE)~\cite{deFlorian:2010aa} and fully resummed (\RHICBOS)~\cite{Nadolsky:2003ga} calculations in Fig.~\ref{Fig:AL}.  The \RHICBOS~calculations are shown for the DSSV08~\cite{deFlorian:2008mr,*deFlorian:2009vb} helicity-dependent PDF set, and the \CHE~calculations are shown for DSSV08~\cite{deFlorian:2008mr,*deFlorian:2009vb} and LSS10~\cite{Leader:2010rb}.  The DSSV08 uncertainties were determined using a Lagrange multiplier method to map out the $\chi^2$ profile of the global fit~\cite{deFlorian:2008mr,*deFlorian:2009vb}, and the $\Delta\chi^2/\chi^2=2\%$ error band in Fig.~\ref{Fig:AL} represents the estimated PDF uncertainty for $A_L^W$~\footnote{DSSV Group, private communication.}. 

The measured $A_L^{\Wpl}$ is negative, consistent with the theoretical predictions.  For $A_L^{\Wmi}$, however, the measured asymmetry is larger than the central value of the theoretical predictions for $\eta_{e^-} < 0$.  This region is most sensitive to the up antiquark polarization, $\Delta\bar{u}$, which is not currently well constrained~\cite{deFlorian:2008mr,*deFlorian:2009vb,Leader:2010rb} as can be seen by the large uncertainty in the theoretical prediction there.  
While consistent within the theoretical uncertainty, the large positive values for $A_L^{\Wmi}$ indicate a preference for a sizable, positive $\Delta\bar{u}$ in the range $0.05~<~x~<~0.2$ relative to the central values of the DSSV08 and LSS10 fits.  Global analyses from both DSSV++~\cite{Aschenauer:2013woa} and neural network PDF~\cite{Nocera:2013yia} have extracted the antiquark polarizations, using our preliminary measurement from the 2012 data set.  These analyses quantitatively confirm the enhancement of $\Delta\bar{u}$ and the expected reduction in the uncertainties of the helicity-dependent PDFs compared to previous fits without our data.

The \Wpm~double-spin asymmetry, shown in Fig.~\ref{Fig:ALL}, is sensitive to the product of quark and antiquark polarizations, and has also been proposed to test positivity constraints using a combination of $A_L$ and $A_{LL}$~\cite{Kang:2011qz}.  The measured double-spin asymmetries are consistent with the theoretical predictions and in conjunction with $A_L^{\Wpm}$ satisfy the positivity bounds within the current uncertainties.

A similar profile likelihood procedure is used to determine the single-spin asymmetry $A_L^{\Zgam}$ for $\Zgam$ production with $|\eta_e|<1.1$, $\EeT>14$ GeV, and $70<\mee<110$ GeV$/c^2$.  $A_L^{\Zgam}$ is sensitive to the combination of $u$, $\bar{u}$, $d$, and $\bar{d}$ polarizations. The measured asymmetry $A_L^{\Zgam} = -0.07^{+0.14}_{-0.14}$ is consistent, within the large uncertainty, with theoretical predictions using the different helicity-dependent PDFs $A_L^{\Zgam}(\textrm{DSSV08}) = -0.07$ and $A_L^{\Zgam}(\textrm{LSS10}) = -0.02$.

In summary, we report new measurements of the parity-violating single-spin asymmetry $A_L$ and parity-conserving double-spin asymmetry $A_{LL}$ for $\Wpm$ production as well as a first measurement of $A_L$ for $\Zgam$ production in longitudinally polarized proton collisions by the \STAR~experiment at \RHIC.  The dependence of $A_L^{\Wpm}$ on the decay lepton pseudorapidity probes the flavor-separated quark and antiquark helicity-dependent PDFs at the $W$ mass scale.  A comparison to theoretical predictions based on different helicity-dependent PDFs suggests a positive up antiquark polarization in the range $0.05~<~x~<~0.2$.  The inclusion of this measurement in global analyses of \RHIC~and \DIS~data should significantly improve the determination of the polarization of up and down antiquarks in the proton and provide new input on the flavor symmetry of the proton's antiquark distributions.

We thank the RHIC Operations Group and RCF at BNL, the NERSC Center at LBNL, the KISTI Center in Korea, and the Open Science Grid consortium for providing resources and support. We are grateful to M. Stratmann for useful discussions.  This work was supported in part by the Offices of NP and HEP within the U.S. DOE Office of Science, the U.S. NSF, CNRS/IN2P3, FAPESP CNPq of Brazil,  the Ministry of Education and Science of the Russian Federation, NNSFC, CAS, MoST and MoE of China, the Korean Research Foundation, GA and MSMT of the Czech Republic, FIAS of Germany, DAE, DST, and CSIR of India, the National Science Centre of Poland, National Research Foundation (NRF-2012004024), the Ministry of Science, Education and Sports of the Republic of Croatia, and RosAtom of Russia.

\bibliography{PRL-W2012-v3.3}

\end{document}